\documentclass[12pt]{article}
\usepackage{cite,amssymb,amsmath,hyperref,authblk}
\begin{document}
\title{\bf New Class of N-dimensional Braneworlds}
\author[1,2]{Merab Gogberashvili \thanks{gogber@gmail.com}}
\author[3]{Pavle Midodashvili \thanks{pmidodashvili@yahoo.com}}
\author[2]{Giorgi Tukhashvili \thanks{gtukhashvili10@gmail.com}}
\affil[1]{Andronikashvili Institute of Physics, 6 Tamarashvili St., Tbilisi 0177, Georgia}
\affil[2] {Javakhishvili State University, 3 Chavchavadze Ave., Tbilisi 0179, Georgia}
\affil[3]{Ilia State University, 3/5 Cholokashvili Ave., Tbilisi 0162, Georgia}
\date{}
\maketitle
\begin{abstract}
The new class of the non-stationary solutions to the system of N-dimensional equations for coupled gravitational and massless scalar field is found. The model represents a single (N-1)-brane in a space-time with one large (infinite) and (N-5) small (compact) space-like extra dimensions. In some particular cases the model corresponds to the gravitational and scalar field standing waves bounded by the brane. These braneworlds can be relevant in string and other higher dimensional models.
\end{abstract}

\vskip 0.5cm

\section{Introduction}

Solutions to the Einstein equations in a higher dimensional space-times for the 3-dimensional singular space-like surfaces with non-factorizable geometry, braneworlds \cite{Hi,brane}, have attracted a lot of interest recently in high energy physics (see \cite{reviews} for reviews). Most of these models were realized as static geometric configurations. However, there have appeared several braneworld models that assumed time-dependent metrics and fields \cite{S}.

A key requirement for realizing the braneworld idea is that the various bulk fields be localized on the brane. For reasons of economy and avoidance of charge universality obstruction \cite{DuRuTi} one would like to have a universal gravitational trapping mechanism for all fields. However, there are difficulties to realize such mechanism with exponentially warped space-times used in standard brane scenario \cite{brane}. In the existing (1+4)-dimensional models spin $0$ and spin $2$ fields are localized on the brane with the decreasing warp factor \cite{brane}, spin $1/2$ field can be localized with the increasing warp factor \cite{BaGa}, and spin $1$ fields are not localized at all \cite{Po}. For the case of (1+5)-dimensions it was found that spin $0$, spin $1$ and spin $2$ fields are localized on the brane with the decreasing warp factor and spin $1/2$ fields again are localized with the increasing warp factor \cite{Od}. There exist also 6D models with non-exponential warp factors that provide gravitational localization of all kind of bulk fields on the brane \cite{6D-old}, however, these models require introduction of unnatural sources.

It turns out that proposed recently brane model \cite{5D-ghost,5D-real}, with standing waves in the bulk, can provide a natural alternative mechanism for universal gravitational trapping of zero modes of all kinds of matter fields. To clarify the mechanism of localization  let us remind that standing electromagnetic waves, so-called optical lattices, can provide trapping of various particles by scattering, dipole and quadruple forces\cite{Opt}. It is known that the motion of test particles in the field of a gravitational wave is similar to the motion of charged particles in the field of an electromagnetic wave\cite{Ba-Gr}. Thus standing gravitational waves could also provide confinement of matter via quadruple forces. Indeed, the equations of motion of the system of spinless particles in the quadruple approximation has the form\cite{Dix}:
\begin{equation} \label{quad}
\frac{Dp^\mu}{ds}= F^\mu = -\frac 16 J^{\alpha\beta\gamma\delta}D^\mu R_{\alpha\beta\gamma\delta}~,
\end{equation}
where $p^\mu$ is the total momentum of the matter field and $J^{\alpha\beta\gamma\delta}$ is the quadruple moment of the stress-energy tensor for the matter field. The oscillating metric due to gravitational waves should induce a quadruple moment in the matter fields. If the induced quadruple moment is out of phase with the gravitational wave the system energy increases in comparison with the resonant case and the fields/particles will feel a quadruple force, $F^\mu$, which ejects them out of the high curvature region, i.e. it would localize them at the nodes.

In this paper we introduce new general class of N-dimensional braneworlds that can provide universal gravitational trapping of all matter fields on the brane. They are realized as standing wave solutions to the system of Einstein and Klein-Gordon equations in N-dimensional space-time. Some particular cases of the introduced general solutions were considered in the papers: \cite{domain,GMS} (for $N=4$), \cite{5D-ghost} (for $N=5$ with the ghost-like scalar field), \cite{5D-real} (for $N=5$ with the real bulk scalar field) and \cite{6D} (for $N=6$).


\section{The Model}

Our setup consists of a single brane and non self interacting scalar field, $\phi$, in $N$-dimensional space-time with the signature $(+, -,...,-)$. The action of the model has the form:
\begin{equation}\label{action}
S = \int d^Nx \sqrt {|g|} \left( \frac{M^{N-2}}{2}R + \frac \epsilon 2 g^{AB}\partial _A \phi \partial _B\phi  + L_{brane} \right)~,
\end{equation}
where $L_{brane}$ is the brane Lagrangian and $M$ is the fundamental scale, which is related to the N-dimensional Newton constant, $G=1/(8\pi M^{N-2})$. The sign coefficient $\epsilon$ in front of the Lagrangian of $\phi$ takes the values $+1$ and $-1$ for the real and phantom-like bulk scalar fields, respectively. Capital Latin indexes numerate $N$-dimensional coordinates, and we use the units where $c = \hbar = 1$.

Variation of the action (\ref{action}) with respect to $g_{AB}$ leads to the Einstein equations:
\begin{equation}\label{EinsteinEquations2}
R_{AB} - \frac 12 g_{AB}R = \frac {1}{M^{N-2}}(\sigma_{AB} + \epsilon T_{AB})~,
\end{equation}
where the source terms are the energy-momentum tensors of the brane,
\begin{equation} \label{sigma}
\sigma _B^A = M^{N-2}\delta (z)\mathrm{diag} \left[\tau_t,\tau_{x_1},...,\tau _{x_{(N - 3)}},\tau_y,\tau_z\right]~,
\end{equation}
with $\tau_N$ being brane tensions, and of the bulk scalar field,
\begin{equation}\label{ScalarFieldEnergyMomentumTensor}
T_{AB} =  \partial _A\phi \partial _B\phi  - \frac 12 g_{AB} \partial ^C\phi \partial _C\phi~.
\end{equation}
The Einstein equations (\ref{EinsteinEquations2}) can be rewritten in the form:
\begin{equation}\label{EinsteinEquations4}
R_{AB}= \frac {1}{M^{N-2}}\left(\sigma_{AB}-\frac{1}{N-2}g_{AB}\sigma + \epsilon \partial_A \phi \partial_B \phi \right)~,
\end{equation}
where $\sigma = g^{AB}\sigma_{AB}$.

We consider the $N$-dimensional metric {\it ansatz}:
\begin{equation}\label{MetricAnsatzGeneral}
ds^2 = (1 + k|z|)^a e^S \left(dt^2 - dz^2\right) - (1 + k|z|)^b \left[e^V \sum\limits_{i = 1}^{N - 3} dx_i^2  + e^{- (N - 3)V}dy^2 \right]~,
\end{equation}
where the curvature scale $k$ and the exponents $a$ and $b$ are some constants, and the metric functions $S = S(t,|z|)$ and $V = V(t,|z|)$ depend only on time, $t$, and on the modulus of the orthogonal to the brane coordinate, $z$. The determinant of (\ref{MetricAnsatzGeneral}) has the form:
\begin{equation}\label{DeterminantOfMetric}
\sqrt {|g|} = e^S (1 + k|z|)^{\frac{b(N - 2) + 2a}{2}}~.
\end{equation}

By the metric (\ref{MetricAnsatzGeneral}) we want to describe geometry of the $(N-1)$-brane placed at the origin of the large space-like extra dimension $z$. Among the $(N-2)$ spatial coordinates of the brane, three of them, $x_1$, $x_2$ and $y$, denote the ordinary infinite dimensions of our world, while the remaining ones, $x_i$ ($i = 3, ..., N-5$), can be assumed to be compact, curled up to the unobservable sizes for the present energies.

The Klein-Gordon equation for the bulk scalar field, $\phi$, in the background metric (\ref{MetricAnsatzGeneral}) has the form:
\begin{equation}\label{BulkScalarFieldEquation}
\left\{ \partial _z^2 + \frac{b(N - 2) k ~\mathrm{sgn} (z)}{2( 1 + k|z|)} \partial _z - \partial_t^2 - (1 + k|z|)^{a - b} e^S \left[ e^{ - V}\sum \limits_{i = 1}^{N - 3} \partial _{x_i}^2 + e^{(N - 3)V}\partial _y^2 \right]\right\}\phi  = 0~.
\end{equation}

The non-zero components of N-dimensional Ricci tensor corresponding to (\ref{MetricAnsatzGeneral}) are:
\begin{eqnarray}\label{RicciTensorComponents1}
R_{tt} &=& \left(ak + S'\right)\delta (z) + \frac 12\left[ S'' + \frac{b(N - 2)k}{2(1 + k|z|)}S' - \ddot S - \frac{(N - 2)(N - 3)}{2}\dot V^2 \right] + \nonumber \\
&+& \frac{a[b(N - 2) - 2]k^2}{4(1 + k|z|)^2}~,\nonumber\\
R_{tz} &=& \frac{b(N - 2)k~\mathrm{sgn} (z)}{4(1 + k|z|)}\dot S - \frac{(N - 2)(N - 3)~\mathrm{sgn}(z)}{4}\dot VV'~, \nonumber\\
R_{x_1x_1} &=& ... = R_{x_{(N - 3)}x_{(N - 3)}} =  - \left(bk + V'\right)e^{- S + V}\delta (z) +  \\
&+& \frac{e^{- S + V}}{(1 + k|z|)^{a - b}}\left\{ \frac 12\left[\ddot V - V'' - \frac{b(N - 2)k}{2(1 + k|z|)}V' \right] - \frac{b[b(N - 2) - 2]k^2}{4(1 + k|z|)^2} \right\}~ , \nonumber\\
R_{yy} &=& - \left[bk - (N - 3)V'\right]e^{- S - (N - 3)V}\delta (z) - \nonumber\\
&-& \frac{e^{- S - (N - 3)V}}{(1 + k|z|)^{a - b}}\left\{\frac{(N - 3)}{2}\left[\ddot V - V'' - \frac{b(N - 2)k}{2(1 + k|z|)}V' \right] + \frac{b[b(N - 2) - 2]k^2}{4(1 + k|z|)^2} \right\}~, \nonumber \\
R_{zz} &=& - \left\{ \left[ a + b(N - 2)\right]k + S' \right \} \delta (z) +  \nonumber \\
&+& \frac 12\left[ \ddot S - S'' + \frac{b(N - 2)k}{2(1 + k|z|)}S' - \frac{(N - 2)(N - 3)}{2}V'^2 \right] + \frac{Dk^2}{4(1 + k|z|)^2}~, \nonumber
\end{eqnarray}
where overdots and primes denote the derivatives with respect to $t$ and $|z|$, respectively, and to shorten the last expression we have introduced the constant:
\begin{equation}
D = a[b(N - 2) + 2] - b(N - 2)(b - 2)~.
\end{equation}

The Einstein equations (\ref{EinsteinEquations4}) can be split into the system of equations for metric functions:
\begin{eqnarray}\label{SystemOfEquationsForMetricFunction}
\frac 12\left[ S'' + \frac{b(N - 2)k}{2(1 + k|z|)}S' - \ddot S - \frac{(N - 2)(N - 3)}{2}\dot V^2 \right] + \frac{a[b(N - 2) - 2]k^2}{4(1 + k|z|)^2} &=& \epsilon \frac{1}{M^{N-2}}\partial _t\phi^2~, \nonumber\\
\frac{b(N - 2)k~\mathrm {sgn}(z)}{4(1 + k|z|)}\dot S - \frac{(N - 2)(N - 3)~\mathrm {sgn}(z)}{4}\dot VV' &=& \epsilon \frac{\mathrm {sgn}(z)}{M^{N-2}}\partial _t\phi \partial _z\phi~, \nonumber \\
\frac{e^{- S + V}}{(1 + k|z|)^{a - b}}\left\{\frac 12\left[ \ddot V - V'' - \frac{b(N - 2)k}{2(1 + k|z|)V'} \right] - \frac{b[b(N - 2) - 2]k^2}{4(1 + k|z|)^2} \right\} &=& \epsilon \frac{1}{M^{N-2}}\partial_{x_1}\phi^2~, \nonumber \\
&...& \\
\frac{e^{- S + V}}{(1 + k|z|)^{a - b}}\left\{ \frac 12 \left[\ddot V - V'' - \frac{b(N - 2)k}{2(1 + k|z|)}V' \right] - \frac{b[b(N - 2) - 2]k^2}{4( 1 + k|z|)^2} \right\} &=& \epsilon \frac{1}{M^{N-2}}\partial_{x_{(N - 3)}}\phi^2~, \nonumber \\
\frac{e^{- S - (N - 3)V}}{(1 + k|z|)^{a - b}}\left\{ \frac{(N - 3)}{2}\left[V'' - \ddot V + \frac{b(N - 2)k}{2(1 + k|z|)}V' \right] - \frac{b[b(N - 2) - 2]k^2}{4(1 + k|z|)^2} \right\} &=& \epsilon \frac{1}{M^{N-2}}\partial_y\phi^2~, \nonumber \\
\frac 12 \left[\ddot S - S'' + \frac{b(N - 2)k}{2(1 + k|z|)}S' - \frac{(N - 2)(N - 3)}{2}V'^2 \right] + \frac{Dk^2}{4(1 + k|z|)^2} &=& \epsilon \frac{1}{M^{N-2}}\partial_z\phi^2~, \nonumber
\end{eqnarray}
and for the brane energy-momentum tensor:
\begin{eqnarray}\label{SystemOfEquationsForBraneEnergyMomentumTensor}
\left( ak + S'\right)\delta (z) &=& \frac{1}{M^{N-2}}\left( \sigma _{tt} - \frac{1}{N - 2}~g_{tt}\sigma  \right)~, \nonumber\\
- \left( bk + V'\right)e^{- S + V}\delta (z) &=& \frac{1}{M^{N-2}}\left(\sigma _{x_1x_1} - \frac{1}{N - 2}~g_{x_1x_1}\sigma  \right)~, \nonumber\\
  &...&\\
- \left( bk + V' \right)e^{- S + V}\delta (z) &=& \frac{1}{M^{N-2}}\left(\sigma _{x_{(N - 3)}x_{(N - 3)}} - \frac{1}{N - 2}~g_{x_{(N - 3)}x_{(N - 3)}}\sigma \right)~, \nonumber\\
- \left[bk - (N - 3)V'\right]e^{- S - (N - 3)V}\delta (z) &=& \frac{1}{M^{N-2}}\left( \sigma _{yy} - \frac{1}{N - 2}~g_{yy}\sigma \right)~, \nonumber\\
- \left\{ [a + b(N - 2)]k + S' \right\} \delta (z) &=& \frac{1}{M^{N-2}}\left( \sigma _{zz} - \frac{1}{N - 2}~g_{zz}\sigma  \right)~. \nonumber
\end{eqnarray}

In the following sections we present all nontrivial solutions to the system of equations (\ref{BulkScalarFieldEquation}), (\ref{SystemOfEquationsForMetricFunction}) and (\ref{SystemOfEquationsForBraneEnergyMomentumTensor}) for our metric {\it ansatz} (\ref{MetricAnsatzGeneral}).


\section{Solutions with $k=0$}

In this section we assume $k=0$ in (\ref{MetricAnsatzGeneral}) and consider the metric:
\begin{equation}\label{MetricAnsatzGeneral2}
ds^2 = e^S \left(dt^2 - dz^2\right) - \left[e^V \sum\limits_{i = 1}^{N - 3} dx_i^2  + e^{- (N - 3)V}dy^2 \right]~.
\end{equation}
The Ricci tensor of this space-time has the $\delta$-like singularity at $z=0$, since the metric functions $S$ and $V$ are depending on of the modulus of $z$. To smooth this singularity we place the brane at the origin of the extra coordinate $z$. Note that without modulus for $z$ the metric (\ref{MetricAnsatzGeneral}) will correspond to the running wave solutions, considered in \cite{plane} for 4D case.

When $k=0$ there exist three independent nontrivial solutions to the system (\ref{BulkScalarFieldEquation}), (\ref{SystemOfEquationsForMetricFunction}) and (\ref{SystemOfEquationsForBraneEnergyMomentumTensor}).


\subsection{The Oscillating Brane}

We start with the simplest case when the bulk scalar field and the metric function $V$ in (\ref{MetricAnsatzGeneral2}) are equal to zero,
\begin{eqnarray}
\phi &=& 0~, \\ \nonumber
V &=& 0~.
\end{eqnarray}
In this case the solution to the system (\ref{BulkScalarFieldEquation}) and (\ref{SystemOfEquationsForMetricFunction}) is:
\begin{equation}
S = \left[ C_1\sin (\Omega t) + C_2\cos (\Omega t)\right]\left[ C_3\sin (\Omega |z|) + C_4\cos (\Omega |z|) \right]~,
\end{equation}
where $C_i$ ($i=1, 2, 3,4$) and $\Omega$ are some constants. The solution corresponds to the oscillating brane at $|z|=0$ in $N$-dimensional space-time.

Imposing on the metric function $S$ the boundary condition on the brane:
\begin{equation}
S|_{|z| = 0} = 0~,
\end{equation}
from the equations (\ref{SystemOfEquationsForBraneEnergyMomentumTensor}) for the brane tensions we find:
\begin{eqnarray}
\tau_t &=& 0~, \nonumber \\
\tau _{x_1} &=& \tau_{x_2} = ... = \tau_{x_{(N - 3)}} = - S'~, \nonumber\\
\tau_y &=& - S'~,\\
\tau_z &=& 0~.
\end{eqnarray}


\subsection{Gravitational and Phantom-like Scalar field Standing Waves}

Now we consider the case with the ghost-like scalar field, $\phi$, when the metric function $S$ is absent in (\ref{MetricAnsatzGeneral2}),
\begin{eqnarray}
\epsilon &=& -1~, \nonumber \\
S &=& 0~.
\end{eqnarray}
Then the equations (\ref{BulkScalarFieldEquation}) and (\ref{SystemOfEquationsForMetricFunction}) have the following solution:
\begin{eqnarray}
V &=& \left[ C_1\sin (\omega t) + C_2\cos (\omega t) \right]\left[ C_3\sin (\omega |z|) + C_4\cos (\omega |z|) \right] ~, \nonumber\\
\phi &=& \frac 12 \sqrt {M^{N-2}(N - 2)(N - 3)} \left[ C_1\sin ( \omega t) + C_2\cos (\omega t)\right]\left[ C_3\sin (\omega |z|) + C_4\cos (\omega |z|) \right],
\end{eqnarray}
where $C_i$ ($i=1, 2, 3,4$) and $\omega$ are some constants. Imposing on the metric function $V$ the boundary condition:
\begin{equation}
V |_{|z| = 0} = 0~,
\end{equation}
from (\ref{SystemOfEquationsForBraneEnergyMomentumTensor}) we find the brane tensions:
\begin{eqnarray}
\tau_t &=& 0~, \nonumber \\
\tau_{x_1} &=& \tau_{x_2} = ... = \tau_{x_{(N - 3)}} =  V'~, \nonumber \\
\tau_y &=& - (N - 3)V'~, \\
\tau _z &=& 0~. \nonumber
\end{eqnarray}


\subsection{Coupled Gravitational and Phantom-like Scalar Field Waves}

Now let us consider the case with the phantom-like scalar field,
\begin{eqnarray}
\epsilon = -1~,
\end{eqnarray}
when the both metric functions $S$ and $V$ are presented in the metric (\ref{MetricAnsatzGeneral2}). In this case the system (\ref{BulkScalarFieldEquation}) and (\ref{SystemOfEquationsForMetricFunction}) has the standing wave solution of the form:
\begin{eqnarray}\label{Solution6}
V &=& \left[ C_1\sin (\omega t) + C_2\cos (\omega t) \right]\left[ C_3\sin (\omega |z|) + C_4\cos (\omega |z|) \right]~, \nonumber \\
\phi &=& \frac 12 \sqrt {M^{N-2}(N - 2)(N - 3)} \left[ C_1\sin (\omega t) + C_2\cos (\omega t) \right]\left[ C_3\sin ( \omega |z|) + C_4\cos (\omega |z|) \right], \\
S &=& \left[ C_5\sin (\Omega t) + C_6\cos (\Omega t) \right]\left[ C_7\sin ( \Omega |z|) + C_8\cos (\Omega |z|) \right]~,\nonumber
\end{eqnarray}
with $C_i$ ($i=1, 2, 3,...,8$), $\Omega$ and $\omega$ being some constants.

Imposing on the metric functions $S$ and $V$ the boundary conditions on the brane:
\begin{eqnarray}
S|_{|z| = 0} &=& 0~, \nonumber\\
V|_{|z| = 0} &=& 0~,
\end{eqnarray}
form (\ref{SystemOfEquationsForBraneEnergyMomentumTensor}) we find the brane tensions:
\begin{eqnarray}
\tau_t &=&  0~, \nonumber\\
\tau_{x_1} &=& \tau_{x_2} = ... = \tau_{x_{(N - 3)}} = - S' + V'~, \nonumber \\
\tau_y &=& - S' - (N - 3)V'~,\\
\tau _z &=& 0~.\nonumber
\end{eqnarray}
From (\ref{Solution6}) it's clear that there are two different frequencies associated with the metric functions $S$ and $V$ ($\Omega$ and $\omega$, respectively), and that the oscillation frequency of the phantom-like bulk scalar field standing wave coincides with the frequency of $V$.


\section{Solutions with $k \neq 0$}

In general case, when the metric curvature scale $k$ in (\ref{MetricAnsatzGeneral}) is non-zero, the system of equations (\ref{BulkScalarFieldEquation}) and (\ref{SystemOfEquationsForMetricFunction}) is self consistent only for the following values of the exponents $a$ and $b$:
\begin{eqnarray}\label{Solution1}
a &=&  - \frac{N - 3}{N - 2}~, \nonumber\\
b &=& \frac{2}{N - 2}~.
\end{eqnarray}
In this case the metric (\ref{MetricAnsatzGeneral}), together with the singularity at $|z|=0$, where we place the brane, and when $k > 0$, has the horizons at $|z| = -1/k$. At these points some components of Ricci tensor get infinite values, while all gravitational invariants, for example the Ricci scalar,
\begin{equation}\label{RicciScalar}
  R = 2\left( S' + \frac{N - 1}{N - 2}k \right)e^{- S}\delta (z) + \left( 1 + k|z| \right)^{(N - 3)/(N - 2)}e^{- S}\left( S'' - \ddot S \right)~,
\end{equation}
are finite.
It resembles the situation with the Schwarzschild Black Hole, however, in contrast, the determinant of (\ref{MetricAnsatzGeneral}) becomes zero at $|z| = -1/k$. As a result nothing can cross these horizons and for the brane observer the extra space $z$ is effectively finite.

For $k \neq 0$ there also exist three independent nontrivial braneworld solutions.


\subsection{The Domain Wall in $N$ Dimensions}

First of all we want to mention the simplest case without the scalar field and metric functions:
\begin{eqnarray}
\phi &=& 0~, \nonumber \\
V &=& 0~, \\
S &=& 0~, \nonumber
\end{eqnarray}
corresponding to the $N$-dimensional generalization of the known 4D \cite{domain} and 5D \cite{5D-real} domain wall solutions:
\begin{equation}
ds^2 = (1 + k|z|)^{- \frac{N - 3}{N - 2}}\left( dt^2 - dz^2 \right) - (1 + k|z|)^{\frac{2}{N - 2}}\left( \sum\limits_{i = 1}^{N - 3} dx_i^2  + dy^2 \right)~.
\end{equation}
In this case the brane tensions in (\ref{sigma}) are:
\begin{eqnarray}
\tau_{t} &=& - 2k~, \nonumber \\
\tau_{x_1} &=& ... = \tau _{x_{(N - 3)}} = \tau_{y} = - \frac{N - 3}{N - 2}~k~, \\
\tau _{z} &=& 0~. \nonumber
\end{eqnarray}


\subsection{Gravitational and Phantom-like Scalar Field Standing Waves Bounded by the Brane}

Now we consider the case with the phantom-like scalar field and when the metric function $S$ is zero:
\begin{eqnarray}
\epsilon &=& -1~, \nonumber \\
S &=& 0~.
\end{eqnarray}
Then the metric (\ref{MetricAnsatzGeneral}) reduces to:
\begin{eqnarray}\label{MetricAnsatz}
ds^2 = (1 + k|z|)^a \left(dt^2 - dz^2\right) - (1 + k|z|)^b \left[ e^V\sum\limits_{i = 1}^{N - 3}dx_i^2 + e^{- (N - 3)V}dy^2 \right]~,
\end{eqnarray}
and the equations (\ref{BulkScalarFieldEquation}) and (\ref{SystemOfEquationsForMetricFunction}) have the following solution:
\begin{eqnarray}
V &=& A \sin (\omega t)J_0(X )~, \nonumber \\
\phi &=& \frac A2 \sqrt {M^{N-2}(N - 2)(N - 3)} \sin (\omega t)J_0(X)~,
\end{eqnarray}
where $A$ and $\omega$ are some constants, $J_0$  is Bessel function of the first kind, and the argument $X$ is defined as:
\begin{equation}\label{X}
X = \frac{|\omega|}{|k|}(1 + k|z|)~.
\end{equation}
Imposing on the metric function $V$ the boundary condition:
\begin{equation}
V |_{|z| = 0} = 0~,
\end{equation}
the equations (\ref{SystemOfEquationsForBraneEnergyMomentumTensor}) for the brane tensions will give the solution:
\begin{eqnarray}
\tau_t &=& - 2k, \nonumber \\
\tau_{x_1} &=& \tau_{x_2} = ... = \tau_{x_{N - 3}} = -\frac{N - 3}{N - 2} k + V'~, \nonumber \\
\tau_y &=& -\frac{N - 3}{N - 2} k - (N - 3)V'~, \\
\tau_z &=& 0~. \nonumber
\end{eqnarray}


\subsection{Standing Wave Braneworld with the Real Scalar Field}

Finally we introduce most realistic general solution to the equations (\ref{BulkScalarFieldEquation}), (\ref{SystemOfEquationsForMetricFunction}) and (\ref{SystemOfEquationsForBraneEnergyMomentumTensor}) with the real scalar field,
\begin{equation}
\epsilon = +1~,
\end{equation}
and all parameters and functions in the metric (\ref{MetricAnsatzGeneral}) being non-zero. Nontrivial solution to (\ref{BulkScalarFieldEquation}) and (\ref{SystemOfEquationsForMetricFunction}) in this case represents $N$-dimensional version of the standing wave braneworld introduced in \cite{5D-real} and is done by:
\begin{eqnarray}
V &=& \left[ C_1 \sin (\omega t) + C_2\cos (\omega t) \right]\left[ C_3J_0(X) + C_4Y_0(X)\right]~, \nonumber \\
\phi &=& \frac 12 \sqrt {M^{N-2}(N - 2)(N - 3)} \left[ C_1\cos (\omega t) - C_2\sin (\omega t)\right]\left[ C_3J_0(X) + C_4Y_0(X) \right]~, \nonumber \\
S &=& \frac 12 (N - 2)(N - 3)X^2\left\{ C_3^2\left[ J_0(X)^2 + J_1(X)^2 - \frac 1X J_0(X)J_1(X) \right] + \right.  \\
&+& C_4^2 \left[ Y_0(X)^2 + Y_1(X)^2 - \frac 1X Y_0(X)Y_1(X) \right] +  \nonumber\\
&+& \left. C_3C_4 \left[ 2\left[ J_0(X)Y_0(X) + J_1(X )Y_1(X) \right] - \frac 1X \left[ J_0(X)Y_1(X) + J_1(X)Y_0(X ) \right] \right] \right\} + C_5~, \nonumber
\end{eqnarray}
where $C_i$ ($i=1, 2, 3, 4, 5$) and $\omega$ are some constants, $J_0$ ($J_1$) and $Y_0$ ($Y_1$) are Bessel functions of the first and the second kind, respectively, and $X$ is defined by (\ref{X}).

Imposing on the metric functions $S$ and $V$ the boundary conditions on the brane:
\begin{eqnarray}
S|_{|z| = 0} &=& 0~, \nonumber \\
V|_{|z| = 0} &=& 0~,
\end{eqnarray}
the system of equations (\ref{SystemOfEquationsForBraneEnergyMomentumTensor}) for the brane tensions will have the following solution:
\begin{eqnarray}
\tau_t &=& - 2k~, \nonumber \\
\tau_{x_1} &=& \tau _{x_2} = ... = \tau _{x_{(N - 3)}} = -\frac{N - 3}{N - 2}k - S' + V'~, \nonumber \\
\tau_y &=& -\frac{N - 3}{N - 2}k - S' - (N - 3)V'~, \\
\tau_z &=& 0~.
\end{eqnarray}


\section{Conclusions}

In this paper we have presented the new class of solutions to the system of the Einstein and Klein-Gordon equations in N-dimensional space-time. The model represents a single (N-1)-brane in a space-time with one large (infinite) and (N-5) small (compact) space-like extra dimensions. We also have derived and explicitly solved the corresponding junction conditions and have obtained analytical expressions for the tensions of the brane. Some cases of the general solutions, which correspond to the braneworlds  in 5D and 6D  with gravitational and scalar field standing waves bounded by the brane, where already studied in \cite{5D-ghost,5D-real,6D}.

Considered in this paper N-dimensional models can be relevant in string and other higher dimensional theories, since they can provide universal gravitational trapping of all kinds of matter fields on the brane, and realize the natural mechanisms of dimensional reduction of multi-dimensional space-times and isotropization of initially non-symmetrically warped braneworlds \cite{Cosm}. 


\section*{Acknowledgments}

MG was partially supported by the grant of Shota Rustaveli National Science Foundation $\#{\rm DI}/8/6-100/12$. The research of PM was supported by Ilia State University (Georgia).



\begin{thebibliography}{99}

\bibitem{Hi} N. Arkani-Hamed, S. Dimopoulos and G. Dvali,
            Phys. Lett. {\bf B 429} (1998) 263,
            arXiv: hep-ph/9803315; \\
             I. Antoniadis, N. Arkani-Hamed, S. Dimopoulos and G. Dvali,
            Phys. Lett. {\bf B 436} (1998) 257,
            arXiv: hep-ph/9804398.

\bibitem{brane} M. Gogberashvili,
               Int. J. Mod. Phys. {\bf D 11} (2002) 1635,
               arXiv: hep-ph/9812296;
               Mod. Phys. Lett. {\bf A 14} (1999) 2025,
               arXiv: hep-ph/9904383; \\
                L. Randall and R. Sundrum,
               Phys. Rev. Lett. {\bf 83} (1999) 3370,
               arXiv: hep-ph/9905221;
               Phys. Rev. Lett. {\bf 83} (1999) 4690,
               arXiv: hep-th/9906064.

\bibitem{reviews} V.A. Rubakov,
                Phys. Usp. {\bf 44} (2001) 871 (Usp. Fiz. Nauk {\bf 171} (2001) 913);\\
                 D. Langlois,
                Prog. Theor. Phys. Suppl. {\bf 148} (2003) 181,
                arXiv: hep-th/0209261;\\
                 P.D. Mannheim,
                {\it Brane-localized Gravity} (World Scientific, Singapore 2005); \\
                 R. Maartens and K. Koyama,
                Living Rev. Rel. {\bf 13} (2010) 5,
                arXiv: 1004.3962 [hep-th].

\bibitem{S}  M. Gutperle and A. Strominger,
            JHEP {\bf 0204} (2002) 018,
            arXiv: hep-th/0202210; \\
             M. Kruczenski, R.C. Myers and A.W. Peet,
            JHEP {\bf 0205} (2002) 039,
            arXiv: hep-th/0204144; \\
             V.D. Ivashchuk and D. Singleton,
            JHEP {\bf 0410} (2004) 061,
            arXiv: hep-th/0407224; \\
             C.P. Burgess, F. Quevedo, R. Rabadan, G. Tasinato and I. Zavala,
            JCAP {\bf 0402} (2004) 008,
            arXiv: hep-th/0310122.

\bibitem{DuRuTi} S.L. Dubovsky, V.A. Rubakov and P.G. Tinyakov,
                JHEP {\bf 0008} (2000) 041,
                arXiv: hep-ph/0007179.

\bibitem{BaGa} B. Bajc and G. Gabadadze,
              Phys. Lett. {\bf B 474} (2000) 282,
              arXiv: hep-th/9912232.

\bibitem{Po} A. Pomarol,
            Phys. Lett. {\bf B 486} (2000) 153,
            arXiv: hep-ph/9911294.

\bibitem{Od} I. Oda,
            Phys. Rev. {\bf D 62} (2000) 126009,
            arXiv: hep-th/0008012.

\bibitem{6D-old} M. Gogberashvili and P. Midodashvili,
                Phys. Lett. {\bf B 515} (2001) 447,
                arXiv: hep-ph/0005298;
                Europhys. Lett. {\bf 61} (2003) 308,
                arXiv: hep-th/0111132; \\
                 M. Gogberashvili and D. Singleton,
                Phys. Lett. {\bf B 582} (2004) 95,
                arXiv: hep-th/0310048;
                Phys. Rev. {\bf D 69} (2004) 026004,
                arXiv: hep-th/0305241; \\
                 M. Gogberashvili, P. Midodashvili and D. Singleton,
                JHEP {\bf 0708} (2007) 033,
                arXiv: 0706.0676 [hep-th].

\bibitem{5D-ghost} M. Gogberashvili and D. Singleton,
                  Mod. Phys. Lett. {\bf A 25} (2010) 2131,
                  arXiv: 0904.2828 [hep-th]; \\
                   M. Gogberashvili, P. Midodashvili and L. Midodashvili,
                  Phys. Lett. {\bf B 702} (2011) 276,
                  arXiv: 1105.1701 [hep-th];\\
                   M. Gogberashvili, P. Midodashvili and L. Midodashvili,
                  Phys. Lett. {\bf B 707} (2012) 169,
                  arXiv: 1105.1866 [hep-th];\\
                   M. Gogberashvili, P. Midodashvili and L. Midodashvili,
                  Int. J. Mod. Phys. {\bf D 21} (2012) 1250081,
                  arXiv: 1209.3815 [hep-th]; \\
                    M. Gogberashvili,
                  JHEP {\bf 09} (2012) 056,
                  arXiv: 1204.2448 [hep-th]; \\
                   M. Gogberashvili, O. Sakhelashvili and G. Tukhashvili,
                  Mod. Phys. Lett. {\bf A 28} (2013) 1350092,
                  arXiv: 1304.6079 [hep-th].

\bibitem{5D-real} M. Gogberashvili and P. Midodashvili,
                 Adv. High Energy Phys. {\bf 2013} (2013) 873686,
                 arXiv: 1310.1911 [hep-th]; \\
                  M. Gogberashvili and P. Midodashvili,
                 Euro. Phys. Lett. {\bf 104} (2013) 50002,
                 arXiv: 1312.6241 [hep-th].

\bibitem{Opt} W.D. Phillips,
             Rev. Mod. Phys. {\bf 70} (1998) 721; \\
              H.J. Metcalf, and P. van der Straten,
             {\it Laser Cooling and Trapping} (Springer, New York 1999); \\
              N. Moiseyev, M. \v{S}indelka and L.S. Cederbaum,
             Phys. Lett. {\bf A 362} (2007) 215;
             Phys. Rev. {\bf A 74} (2006) 053420.

\bibitem{Ba-Gr} D. Baskaran and L.P. Grishchuk,
               Class. Quant. Grav. {\bf 21} (2004) 4041,
               arXiv: gr-qc/0309058.

\bibitem{Dix} W.G. Dixon,
             Nuovo Cim. {\bf 34} (1964) 318;
             Proc. R. Soc. London {\bf A 314} (1970) 499;
             Gen. Rel. Grav. {\bf 4} (1973) 199.

\bibitem{domain} A. H. Taub,
                Phys. Rev. {\bf 103} (1956) 454; \\
                 A. Vilenkin,
                Phys. Rev. {\bf D 23} (1981) 852; \\
                 J. Ipser and P. Sikivie,
                Phys. Rev. {\bf D 30} (1984).

\bibitem{GMS} M. Gogberashvili, S. Myrzakul and D. Singleton,
             Phys. Rev. {\bf D 80} (2009) 024040.
             arXiv: 0904.1851 [gr-qc].

\bibitem{6D} L.J.S. Sousa, J.E.G. Silva and C.A.S. Almeida,
            arXiv: 1209.2727 [hep-th];\\
             L.J.S. Sousa, J.E.G. Silva and C.A.S. Almeida,
            Phys. Rev. {\bf D 89} (2014),
            arXiv: 1311.5848 [hep-th]; \\
             P. Midodashvili,
            Int. J. Theor. Phys. {\bf 53} (2014),
            arXiv: 1211.0206 [hep-th];\\
             O. Sakhelashvili,
            Int. J. Theor. Phys. {\bf 53} (2014),
            arXiv: 1311.1030 [gr-qc].

\bibitem{plane} U. Yurtsever,
               Phys. Rev. {\bf D 38} (1988) 1706; \\
                A. Feinstein and J. Iba\~{n}ez,
               Phys. Rev. {\bf D 39} (1989) 470; \\
                J. Griffiths,
               {\it Colliding Plane Waves in General Relativity}
               (Oxford University Press, Oxford 1991), Chapter 10.

\bibitem{Cosm} M. Gogberashvili, A. Herrera-Aguilar, D. Malag\'on-Morej\'on, R.R. Mora-Luna and U. Nucamendi,
              Phys. Rev. {\bf D 87} (2013) 084059,
              arXiv: 1201.4569 [hep-th]; \\
               M. Gogberashvili, A. Herrera-Aguilar, D. Malag\'on-Morej\'on and R.R. Mora-Luna,
              Phys. Lett. {\bf B 725} (2013) 208,
              arXiv: 1202.1608 [hep-th].

\end{thebibliography}
\end{document}